\documentclass[prd,nofootinbib,twocolumn,superscriptaddress,preprintnumbers,
               balancelastpage,floatfix,showkeys,10pt,aps]{revtex4-1}
\usepackage{graphicx,amsmath,amssymb,hyperref,natbib,slashed}
\usepackage[dvipsnames]{xcolor}
\usepackage{xspace}
\usepackage{bbm} 

\usepackage[utf8]{inputenc}
\usepackage{placeins}

\begin{document}
\title{Probing the freeze-in mechanism in dark matter models with $U(1)'$ gauge extensions}

\preprint{TTK-19-33}
 
  \author{Saniya Heeba}
 \email{heeba@physik.rwth-aachen.de}

 \author{Felix Kahlhoefer}
 \email{kahlhoefer@physik.rwth-aachen.de}
 
 \affiliation{Institute for Theoretical Particle Physics and Cosmology (TTK), RWTH Aachen University, D-52056 Aachen, Germany}

\begin{abstract}

New gauge bosons at the MeV scale with tiny gauge couplings (so-called dark photons) can be responsible for the freeze-in production of dark matter and provide a clear target for present and future experiments. 
We study the effects of thermal mixing between dark photons and Standard Model gauge bosons and of the resulting plasmon decays on dark matter production before and after the electroweak phase transition. 
In the parameter regions preferred by the observed dark matter relic abundance, the dark photon is sufficiently long-lived to be probed with fixed-target experiments and light enough to induce direct detection signals. 
Indeed, current limits from XENON1T already constrain Dirac fermion dark matter in the GeV to TeV range produced via the freeze-in mechanism.
We illustrate our findings for the case of a $U(1)_{B-L}$ gauge extension and discuss possible generalisations.

\end{abstract}

\vspace{2mm}

\maketitle


\section{Introduction}

Given our complete ignorance of the nature of dark matter (DM), it is sensible to construct DM models in analogy to phenomena known from the Standard Model (SM). Following this approach, an attractive possibility is that DM particles carry a charge $g_\mathrm{DM}$ under a new $U(1)'$ gauge group and have interactions similar to the ones between photons and electrons. Particular attention has been paid to the case that the SM is neutral under this new gauge group and that tiny couplings are induced via a kinetic mixing parameter $\epsilon$. It has been shown that for $\epsilon \, g_\mathrm{DM} \sim 10^{-11}$ it is possible to reproduce the observed DM relic abundance via the freeze-in mechanism~\cite{Hall:2009bx,Chu:2011be,Chu:2013jja,Bernal:2017kxu}. 

Intriguingly, in spite of the small coupling, it may be possible to test such a scenario with direct detection experiments, provided the mediator (the so-called dark photon) is sufficiently light, such that scattering cross sections are enhanced~\cite{Essig:2011nj,Essig:2015cda,Battaglieri:2017aum,Hambye:2018dpi}. Indeed, much of the literature has focused on the case that the mediator is either exactly massless or that its mass can be neglected in all calculations. In this case, there is a one-to-one correspondence between the relic density requirement and direct detection signatures.

In the present work, we consider the alternative possibility that the mediator has small couplings to both the DM particle and the SM. The simplest realisation of this idea is that both the DM particle as well as SM fermions are charged under a new $U(1)'$ gauge group, for example $U(1)_{B-L}$ (see e.g. Ref.~\cite{Knapen:2017xzo}). The cross section of any process leading to the production of DM particles is then proportional to the fourth power of the new gauge coupling $g'$, such that values of the order $g' \sim 10^{-6}$ lead to the observed DM relic abundance~\cite{Kaneta:2016vkq}. 

Our central observation is that such a scenario differs in a number of important ways from the case of a dark photon with tiny kinetic mixing. First of all, for gauge couplings of the required magnitude, the new gauge boson $A^\prime$ enters into thermal equilibrium with the SM thermal bath through processes like $q g \to q A'$~\cite{Evans:2017kti}. As a result, the density of $A^\prime$ bosons is given by an equilibrium distribution, opening up a new way to produce DM particles: $A^\prime A^\prime \to \chi \chi$. This channel depends in a different way on the specific $U(1)'$ charge assignments than processes involving both DM and SM particles and therefore opens up additional parameter space in the translation between the relic density requirement and direct detection experiments.

Furthermore, if the $A^\prime$ boson is in thermal equilibrium, its mass must be larger than a few MeV in order to satisfy constraints on the number of relativistic degrees of freedom during Big Bang Nucleosynthesis (BBN)~\cite{Cyburt:2015mya} and on exotic sources of energy injection~\cite{Hufnagel:2018bjp}. In this mass range and for the required coupling strength, the dark photons are long-lived and can be searched for in a number of ways at the intensity frontier~\cite{Beacham:2019nyx}, using for example fixed target experiments like SeaQuest~\cite{Berlin:2018pwi} and SHiP~\cite{Alekhin:2015byh} or forward detectors like Faser~\cite{Feng:2017uoz} (see Ref.~\cite{Bauer:2018onh} for a recent sensitivity study and Ref.~\cite{Belanger:2018sti} for other ways to probe freeze-in at colliders).

At first sight, the fact that the mediator needs to be rather heavy reduces the prospects of testing these models with direct detection experiments. If the DM mass is at the GeV to TeV scale, however, the mediator mass can be comparable to the typical momentum transfer (which is of the order of $10^{-3} m_\chi$) such that direct detection signals may be observable~\cite{Fornengo:2011sz,Kahlhoefer:2017ddj}. We perform a detailed calculation of the resulting event rates in direct detection experiments and show that existing experiments already possess some sensitivity to our model, while future experiments will be able to explore broader regions of parameter space, which are difficult to access with accelerator experiments.

We also point out a number of subtleties relevant for the calculation of the DM abundance from freeze-in. Whenever the thermal plasma contains particles charged under several $U(1)$ gauge groups, plasma effects induce mass mixing between the corresponding gauge bosons~\cite{Comelli:1996vm}. As a result, both the hypercharge gauge boson (before electroweak symmetry breaking, EWSB) and the photon (after EWSB) mix with the $A'$ boson, leading to two mass eigenstates that can mediate processes involving DM particles. This finding is particularly relevant if the thermal mass of one of these eigenstates is large enough to allow for direct decays into DM particles. Such plasmon decays have recently been found to give a relevant contribution to the freeze-in production of sub-MeV DM~\cite{An:2018nvz,Dvorkin:2019zdi}, and we extend the discussion there to more general gauge groups and to temperatures above the electroweak phase transition (EWPT).

This paper is structured as follows. In section~\ref{sec:model}, we present the general model set-up and discuss thermal effects with a particular focus on the induced mass mixing between gauge bosons. A detailed calculation of the DM relic density via the freeze-in mechanism is provided in section~\ref{sec:relic}, where we also discuss the differences before and after EWSB. Finally, section~\ref{sec:constraints} deals with experimental constraints on our model from direct detection experiments. Our conclusions are presented in section~\ref{sec:conclusions}.

\section{General set-up}
\label{sec:model}

In the present work we consider $U(1)^\prime$ gauge extensions of the SM, which give rise to a new gauge boson $A^\prime$ (also called \emph{dark photon}). Further, we introduce a Dirac fermion $\chi$ that is charged under the new gauge group and plays the role of DM in our model. The relevant part of the vacuum Lagrangian is then given by:
\begin{align}
\mathcal{L} \supset  \frac{1}{2}m_{A^\prime}^2 A^\prime_\mu A^{\prime \mu} - g_\mathrm{DM}A^\prime_\mu \bar{\chi}\gamma^\mu\chi - \sum\limits_f g^\prime q^\prime_f A^\prime_\mu\bar{f}\gamma^\mu f
\label{eq:L}
\end{align}
where the sum runs over all SM fermions with $U(1)^\prime$ charge $q^\prime_f$. In eq.~(\ref{eq:L}) we have introduced the effective coupling $g_\mathrm{DM}$, which is given by the product of the $U(1)^\prime$ gauge coupling $g^\prime$ and the DM charge $q^\prime_\chi$. 

For concreteness, we will focus on one of the simplest gauge extensions of the SM, namely gauged baryon number minus lepton number, $B-L$. This model can be made anomaly-free by adding three massive right-handed neutrinos (RHNs), which we assume to be sufficiently heavy to have no effect on the DM phenomenology.\footnote{Likewise, we assume that the scalar field that gives mass to the RHNs and the $A^\prime$ has a very large mass and can be ignored in the present context. The case where the right-handed neutrinos themselves are DM candidates produced via the freeze-in mechanism is discussed in Ref.~\cite{Biswas:2016bfo}.} The $U(1)^\prime$ charges of the SM fermions are then given by $q'_f = 1/3$ for quarks and $q'_f = -1$ for leptons and the gauge coupling is denoted by $g_\text{BL}$. In the interest of generality we will not specify the charge $q^\prime_\chi$ of the DM particle and instead treat $g_\text{DM}$ as a free parameter. It is, however, worth emphasising that for $q_\chi' \neq -1$, the stability of DM is automatically guaranteed by the $U(1)_{B-L}$ gauge symmetry.
It will be convenient to define the ratio of the two couplings as
\begin{align}
r \equiv \frac{g_\mathrm{BL}}{g_\mathrm{DM}} = \frac{1}{q_\chi'}\,.
\end{align}
Thus, the free parameters in our model are $m_\chi,\,m_{A^\prime}, \,g_\mathrm{BL},$ and $g_\mathrm{DM}$.

The model introduced above can, in principle, account for the observed DM relic abundance for a wide range of DM masses via several different mechanisms. Here, we will be most interested in DM production via the freeze-in mechanism, which requires $g_\mathrm{BL}, g_\mathrm{DM} \ll 1$, and constraints from direct detection experiments, which are most relevant for DM masses in the GeV range. While the DM particles are assumed to have a negligible initial abundance and never enter into thermal equilibrium, we will see below that the dark photon does in general thermalise with the bath of SM particles. We will therefore limit ourselves to $m_{A^\prime} \gtrsim 1 \, \mathrm{MeV}$, such that constraints from BBN are readily satisfied. However, before discussing the thermal history of the dark sector in detail, we first need to discuss how the Lagrangian in eq.~(\ref{eq:L}) is modified at finite temperatures and densities.

\subsection{Temperature-induced mixing}

The generation of a mass term for the photon due to its interactions in a QED plasma is a well-known result of stellar physics~\cite{Braaten:1993jw,Raffelt:1996wa}. In the presence of an additional $U(1)^\prime$ gauge group, similar interactions also induce a mixing between the two $U(1)$ gauge bosons. The importance of this mixing is elaborated upon in recent works in the context of non-thermal production of light DM for the case of a plasma composed primarily of electrons~\cite{Knapen:2017xzo,Dvorkin:2019zdi}. For heavier DM particles and higher temperatures, this assumption is inaccurate and a more general formalism is needed.

With that in mind, let us consider the Lagrangian
\begin{equation}
 \mathcal{L} \supset - \sum_f e \, q_f \bar{f} \gamma^\mu f \, A_\mu + g' \, q'_f \bar{f} \gamma^\mu f \, A'_\mu \; ,
\end{equation}
where $A$ denotes the SM photon and $q_f$ is the electromagnetic charge of the SM fermion $f$. We assume that the $U(1)^\prime$ gauge coupling satisfies $g' \ll e$ and that the dark charges $q'_f$ are of order unity.

In a thermal bath with temperature $T$ the photon mass is given to first approximation by the plasma frequency~\cite{Raffelt:1996wa}:
\begin{equation}
\label{eq:PlasmonMass}
 m_A^2 \approx \omega_P^2 \approx \sum_f \frac{e^2 \, q_f^2}{9} T^2 \equiv q_\text{eff} \frac{e^2 \, T^2}{9}\; ,
\end{equation}
where the sum includes \emph{only relativistic} fermions and $q_\text{eff}$ denotes the effective number of charge degrees of freedom.\footnote{The formula assumes four degrees of freedom per fermion, i.e.\ it includes the anti-particle with opposite charge. Also, we implicitly assume that the photon itself is non-relativistic. For highly relativistic photons, the thermal mass is larger by a factor 3/2.} In principle, the dark photon also obtains a thermal correction to its mass, which is given by
\begin{equation}
\Delta m_{A'}^2 \approx \sum_f \frac{g'^2 \, q'^2_f}{9} T^2 \; .
\end{equation}
For small $g'$ this contribution is however negligible.

More importantly, plasma effects will induce a mass mixing of the form $\delta m^2 A^\mu A'_\mu$ with~\cite{Comelli:1996vm}
\begin{equation}
 \delta m^2 \approx \sum_f \frac{e\,g' \, q_f \, q'_f}{9} T^2 \equiv q'_\text{eff} \frac{e \, g' \, T^2}{9}\; . 
\end{equation}
In contrast to $q_\text{eff}$, the effective mixing degrees of freedom $q'_\text{eff}$ do not necessarily increase monotonically with temperature, as there can be negative contributions if $q_f$ and $q'_f$ have opposite sign. Since by assumption $g' \ll e$ one finds $\delta m^2 \ll m_A^2$.

To remove the mass mixing, one needs to make the replacements $A \to A - \theta A'$ and $A' \to A' + \theta A$ with
\begin{equation}
 \theta = \frac{\delta m^2}{m_A^2 - m_{A'}^2} \; .
\end{equation}
If $m_{A'} \ll m_A$, this expression simplifies to
\begin{equation}
\label{eq:mix_after}
 \theta = \frac{g'}{e} \frac{q'_\text{eff}}{q_\text{eff}} \; .
\end{equation}
The Lagrangian from above then becomes
\begin{align}
 \mathcal{L} & \supset - \sum_f e \, q_f \bar{f} \gamma^\mu f \, (A_\mu - \theta A'_\mu) + g' \, q'_f \bar{f} \gamma^\mu f \, (A'_\mu + \theta A) \nonumber \\
 & = - \sum_f (e q_f + \theta g' q'_f) \bar{f} \gamma^\mu f A_\mu  + (g' q'_f - \theta e  q_f) \bar{f} \gamma^\mu f A'_\mu \, .
\end{align}

If only electrons contribute to $q_\text{eff}$ and $q'_\text{eff}$ one finds $\theta = g' \, q'_e / (e \, q_e)$, such that the dark photon coupling to electrons vanishes. The same is true if several fermions contribute to the effective number of charge degrees of freedom, provided all dark charges are proportional to the electromagnetic charges, as for example in models with kinetic mixing.\footnote{In fact this feature is precisely what prevents a kinetically-mixed dark photon with tiny mass from thermalising with the SM thermal bath and makes it possible to satisfy constraints on the number of relativistic degrees of freedom~\cite{Knapen:2017xzo,Dvorkin:2019zdi}.} In general, however, the ratio of the effective charge and mixing degrees of freedom will be a complicated function of temperature. 

Let us now return to the case of a $B-L$ gauge boson. Each charged lepton contributes 1 to both $q_\text{eff}$ and $q'_\text{eff}$, each up-type quark contributes $\tfrac{4}{3}$ to $q_\text{eff}$ but only $\tfrac{2}{3}$ to $q'_\text{eff}$ (including colour degrees of freedom) and each down-type quark contributes $\tfrac{1}{3}$ to $q_\text{eff}$ but $-\tfrac{1}{3}$ to $q'_\text{eff}$ (see Table \ref{tab:chargeafterEW}). At temperatures above the bottom-quark mass and below the top-quark mass one therefore finds $q_\text{eff} = 3 + 2 \times \tfrac{4}{3} + 3 \times \tfrac{1}{3} = \tfrac{20}{3}$ and $q'_\text{eff} = 3 + 2 \times \tfrac{2}{3} - 3 \times \tfrac{1}{3} = \tfrac{10}{3}$. Hence, we find
\begin{align}
\theta_\mathrm{BL} = \frac{1}{2}\frac{g_\mathrm{BL}}{e}\,,
\end{align}
which implies that the dark photon will not couple to fermions with $q_f = 2 q'_f$ (i.e.\ up-type quarks), but it will couple to all other SM fermions.

\begin{table}
\begin{center}
\begin{tabular}{ c c c c c }
\hline
\hline
Particle & $N_c$ & $N_f$ & $q_f$ & $B-L$ \\ \hline
$u$ & 3 & 3 & $2/3$ & $1/3$ \\ 
$d$ & 3 & 3 & $-1/3$ & $1/3$ \\ 
$e$ & 1 & 3 & $-1$ & $-1$ \\ 
$\nu_\mathrm{L}$ & 1 & 3 & $0$ & $-1$ \\ 
$\nu_\mathrm{R}$ & 1 & 3 & $0$ & $-1$ \\ 
\hline
\hline
\end{tabular}
\end{center}
\vspace{-2mm}
\caption{\label{tab:chargeafterEW}$B-L$ and charge assignments after EWSB.}
\vspace{2mm}
\begin{center}
\begin{tabular}{ c c c c c}
\hline\hline
Particle & $N_c$ & $N_f$ & $Y$ & $B-L$ \\ \hline
$q_\mathrm{L}$ & 3 & 3 & $1/6$ & $1/3$ \\ \hline
$u_\mathrm{R}$ & 3 & 3 & $2/3$ & $1/3$ \\ \hline 
$d_\mathrm{R}$ & 3 & 3 & $-1/3$ & $1/3$ \\ \hline
$e_\mathrm{L}$ & 1 & 3 & $-1/2$ & $-1$ \\ \hline
$e_\mathrm{R}$ & 1 & 3 & $-1$ & $-1$ \\ \hline
$\nu_\mathrm{R}$ & 1 & 3 & $0$ & $-1$ \\ \hline
$H$ & 1 & 1 & $1/2$ & $0$ \\ \hline\hline
\end{tabular}
\end{center}
\vspace{-2mm}
\caption{\label{tab:chargebeforeEW}$B-L$ and hypercharge assignments before EWSB.}
\end{table}

\begin{figure*}
\includegraphics[width=0.4\textwidth,clip,trim={0 0 0 0}]{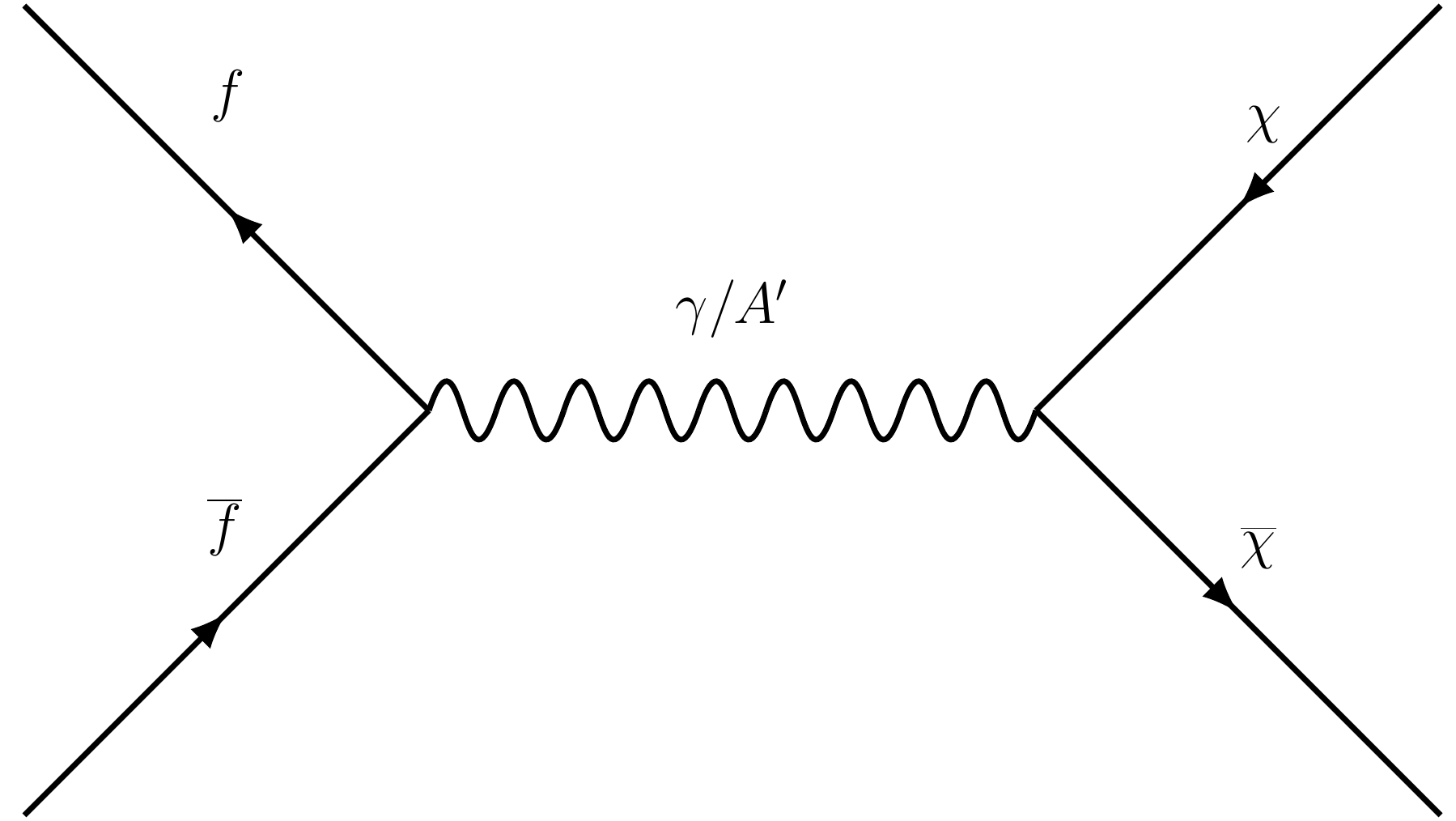}\qquad
\includegraphics[width=0.32\textwidth]{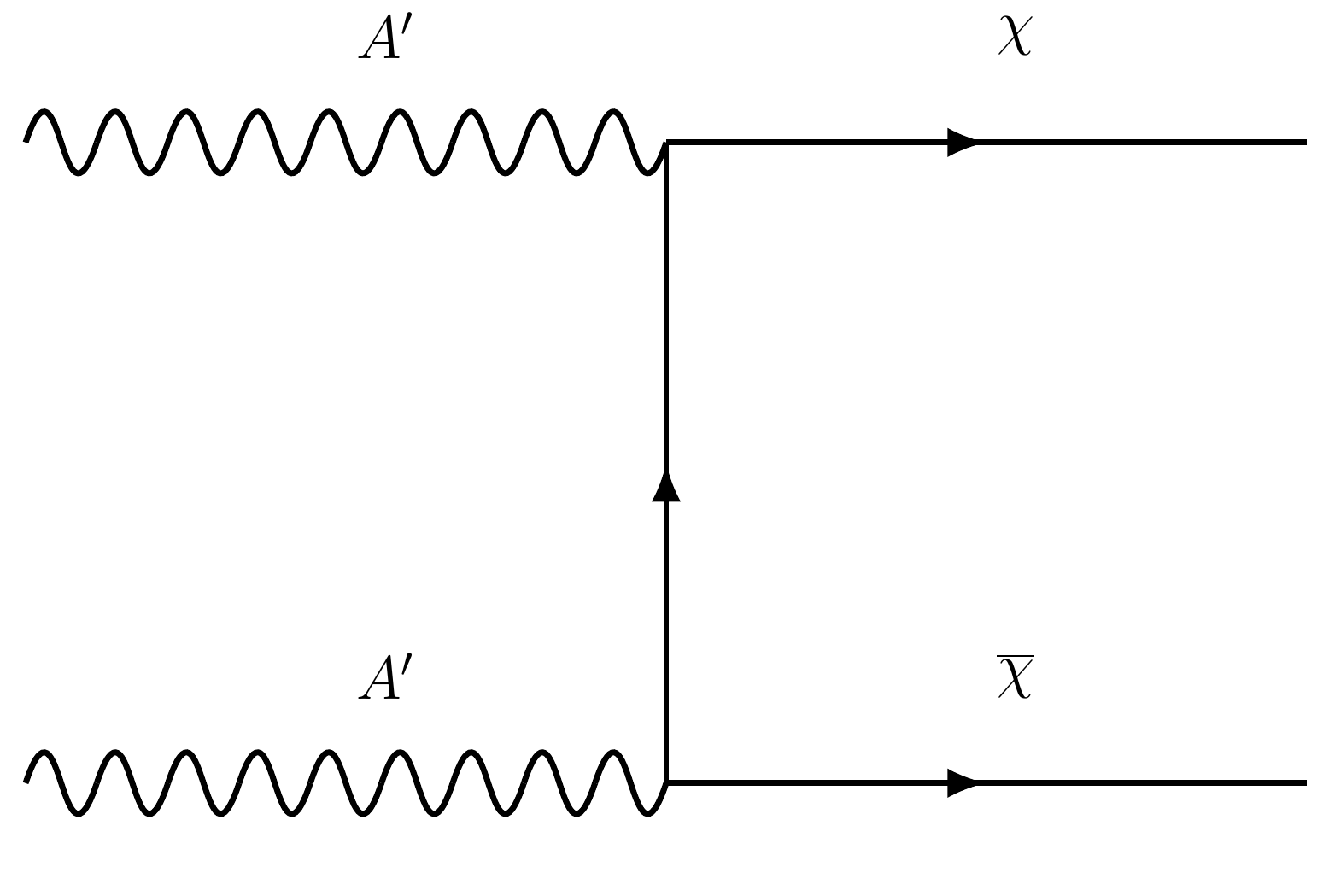}
\caption{\label{fig:prod_channels} Production channels for freeze-in. Left: production from the annihilation of SM fermions $f$ and $\bar{f}$. Right: production from the annihilation of dark photons $A^\prime$.} 
\end{figure*}

A completely analogous reasoning can be applied to mixing with the $U(1)_Y$ hypercharge gauge boson before EWSB. In this case, the effective charge degrees of freedom are defined as~\cite{Rychkov:2007uq}
\begin{equation}
 q_\text{eff} = \sum_{f_L} Y_{f_L}^2 + \frac{1}{2} \sum_{f_R} Y_{f_R}^2 + 2 Y_{H}^2
\end{equation}
and analogously for $q'_\text{eff}$. Hence, using Table \ref{tab:chargebeforeEW} for temperatures above the EWPT, one finds $q_\text{eff} = 3 \times (\tfrac{1}{12} + \tfrac{1}{4} + \tfrac{2}{3} + \tfrac{1}{6} + \tfrac{1}{2}) + 2 \times \tfrac{1}{4} = \tfrac{11}{2}$ and (for a $B-L$ gauge boson) $q'_\text{eff} = 4$.

The thermal mass of the hypercharge gauge boson before EWSB is then given by
\begin{equation}
m_Y^2 = q_\text{eff} \frac{g_Y^2 \,T^2}{9} = \frac{11}{2} \frac{g_Y^2 T^2}{9}\,,
\end{equation}
where $g_Y$ is the hypercharge gauge coupling\footnote{We assume that the running of this coupling with temperature is small enough to be ignored.}, and the mixing is defined by
\begin{equation}
\delta m^2 = q'_\text{eff} \frac{g_Y\,g_\mathrm{BL}\,T^2}{9} = 4 \frac{g_Y\,g_\mathrm{BL}\,T^2}{9}\, ,
\end{equation}
such that
\begin{equation}
 \theta_\mathrm{BL} = \frac{8}{11} \frac{g_\mathrm{BL}}{g_Y}\,.
\end{equation}

To conclude this discussion, we note that there is no mixing between $W_3$ and $A^\prime$ before EWSB~\cite{Comelli:1996vm}. After EWSB, however, temperature effects can induce a mixing between the $Z$ boson and the dark photon. By following a similar line of reasoning as above, one can show that this mixing is suppressed by the mass of the Z-boson and can typically be neglected. 

\subsection{Temperature-induced masses}

Just as interactions in a thermal plasma impart a mass to the photon, they also result in corrections to the mass terms of other bath particles. These corrections are sub-leading at low temperatures but become increasingly important as the temperature increases. Before the EWPT, these corrections are what make the fermions massive. We divide the thermal history of the universe in two regimes: before and after the phase transition.

Before the EWPT, the fermions are chiral with left- and right-handed fermions belonging to different representations of the SM gauge group. As a result, the temperature-induced masses of these fermions are also unequal and given by \cite{Elmfors:1993re},
\begin{align}
m_{l,\mathrm{L}}^2 &= \frac{m_Z^2 + 2 m_W^2 + m_l^2 + m_{l^\prime}^2}{2 v_h^2}T^2\,,\\
m_{l,  \mathrm{R}}^2 &= \frac{m_Z^2 - m_W^2 + \frac{1}{2}m_l^2}{2v_h^2} T^2\,,
\end{align}
where $m_{l,l^\prime}$ are the zero-temperature masses of the leptons belonging to the same $SU(2)$ doublet, $v_h$ is the Higgs vacuum expectation value, and $m_Z$ and $m_W$ denote the zero-temperature masses of the $Z$ and $W$ bosons. In particular, the masses of left-handed fermions belonging to the same $SU(2)$ doublet are equal. Similarly, for first-generation quarks one finds,
\begin{align}
m_{q,\mathrm{L}}^2 &= \frac{1}{6} g_s^2 T^2 + \frac{3m_W^2 + \frac{1}{9}(m_Z^2 - m_W^2) + m_{u}^2 + m_{d}^2}{8 v_h^2}T^2\,,\\
m_{u, \mathrm{R}}^2 &= \frac{1}{6} g_s^2 T^2 + \frac{\frac{4}{9}(m_Z^2 - m_W^2) + \frac{1}{2}m_{u}^2}{2 v_h^2}T^2\,,\\
m_{d,\mathrm{R}}^2 &= \frac{1}{6} g_s^2 T^2 + \frac{\frac{1}{9}(m_Z^2 - m_W^2) + \frac{1}{2}m_{d}^2}{2 v_h^2}T^2\,,
\end{align}
where $m_{u, d}$ are the zero-temperature masses of up and down quarks and $g_s$ is the strong-coupling constant. The mass terms for the second and third generation are given in complete analogy.

After the EWPT, the mass corrections are \cite{Elmfors:1993re},
\begin{align}
\Delta m_l^2 = \frac{e^2\,T^2}{8}\,,\\
\Delta m_q^2 = \frac{g_s^2 T^2}{6}\,. 
\end{align} 

In summary, we have shown that in-medium effects do not only alter the masses of the various particles, but also induce mixing between different $U(1)$ gauge groups, which may significantly change the coupling structure of the model at high temperatures. Since these effects differ before and after the EWPT, it is important to treat DM production in these two temperature regimes separately. The resulting subtleties will be the focus of the following section. 

\section{Freezing-in dark matter}
\label{sec:relic}

\begin{figure*}
\centering
\includegraphics[width=0.98\textwidth]{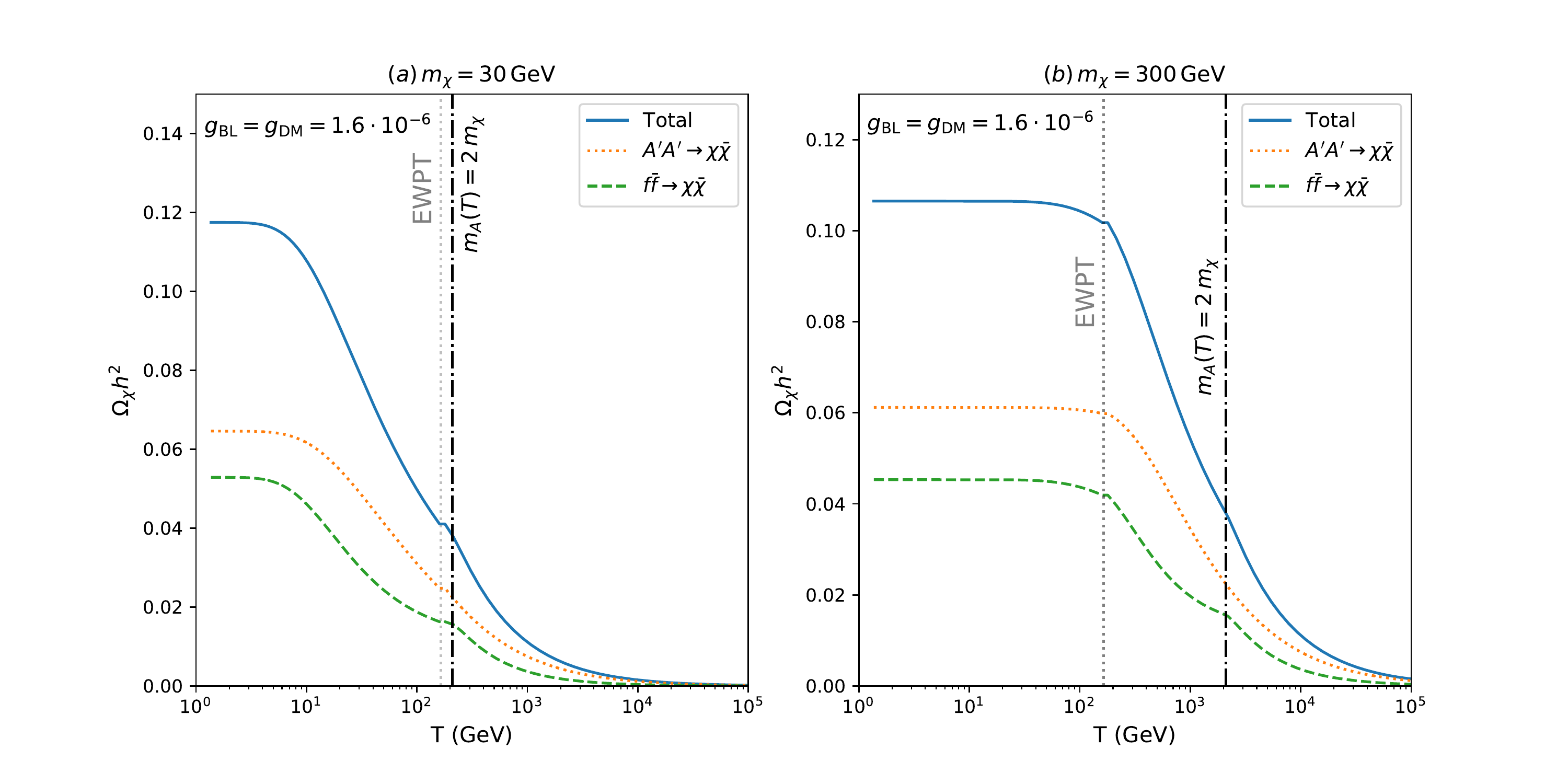}
\caption{\label{fig:Abundance_temp} The DM abundance $\Omega_\chi h^2$ defined in eq.~(\ref{eq:Omega}) as a function of temperature $T$ for two values of the DM mass $m_\chi$. Vertical lines indicate the temperature of the EWPT and the temperature for which hypercharge plasmon decays become kinematically forbidden.}
\end{figure*}

As already mentioned, we are interested in the case where the initial abundance of DM is negligible and the dark sector is slowly populated by the decays and annihilations of particles in the thermal bath, which we assume to be comprised of both SM particles and dark photons. For the dark photons to enter into thermal equilibrium with the SM, we require that the interaction rate of processes like $q g \to q A^\prime$ are larger than the Hubble rate $H$ for $T \approx m_\chi$, which for $m_\chi < 1 \, \mathrm{TeV}$ is always the case for $g_\text{BL} \gtrsim 10^{-7}$~\cite{Evans:2017kti}. At the same time, we assume that $g_\text{DM}$ is sufficiently small that the DM particle does not enter into thermal equilibrium.

Under these assumptions, DM can be created via the freeze-in mechanism by two kinds of processes differentiated by their dependence on couplings (see Fig. \ref{fig:prod_channels}): 

\textbf{a) Annihilation or decay of SM particles }: Processes of the form $f\bar{f} \to \chi\bar{\chi}$ have cross-sections that scale as $g_\mathrm{BL}^2 g_\text{DM}^2$. This is true not only for the direct contribution from a dark photon in the s-channel but also for the (non-negligible) contribution arising from the thermal mixing between the dark and the visible photon. For $m_A(T) > 2 m_\chi$, DM particles can also be produced via the decay $A\to\chi\bar{\chi}$. The partial decay width for this process is again induced by thermal mixing and therefore scales proportional to $g_\text{DM}^2 \, g_\text{BL}^2$.

\textbf{b) Annihilation of dark photons}: The cross section for the process $A^\prime A^\prime \to \chi \bar{\chi}$ scales as $g_\text{DM}^4$. Consequently, this production channel dominates for $g_\text{DM} \gg g_\mathrm{BL}$, while direct production from SM particles gives the dominant contribution if $g_\mathrm{BL} \gg g_\text{DM}$.

The freeze-in abundance can then be calculated by solving the standard Boltzmann Equation~\cite{Dvorkin:2019zdi},
\begin{align}
\frac{s}{2} \frac{\mathrm{d}{Y}_\chi}{\mathrm{d}t} = \langle \sigma_{f\bar{f}} v \rangle n^2_f + \langle \sigma_{A^\prime A^\prime} v \rangle n^2_{A^\prime} + \langle \Gamma_{A \to \chi \bar{\chi}}\rangle n_A\,,
\end{align} 
where $Y_\chi \equiv n_\chi / s$ and $s$ denotes the entropy density. The final term on the right-hand side contributes only for $m_A(T) > 2 m_\chi$, corresponding to $T \gtrsim 7 m_\chi$, while the first two terms remain efficient until $T \approx m_\chi$ and therefore typically dominate the total yield. For $T \gg m_\chi$ the thermally-averaged cross sections $\langle \sigma_{f\bar{f}} v \rangle$ and $\langle \sigma_{A^\prime A^\prime} v \rangle$ are proportional to $g_\text{BL}^2 g_\text{DM}^2 / T^2$ and $g_\text{DM}^4 / T^2$, respectively. The corresponding contributions to the total yield then scale as
\begin{align}
 Y_{f\bar{f}} \propto \frac{g_\text{BL}^2 \, g_\text{DM}^2 \, M_\text{P}}{m_\chi} \, ,&   & Y_{A^\prime A^\prime} \propto \frac{g_\text{DM}^4 \, M_\text{P}}{m_\chi} \; ,
\end{align}
where $M_\text{P}$ denotes the Planck mass and we have neglected a slight dependence on the DM mass of the number of relativistic degrees of freedom $g_\ast$. In this approximation, the DM abundance, which is given by
\begin{align}
\Omega_\chi  = \frac{m_\chi\, Y_\chi s_0}{\rho_{c,0}}
\label{eq:Omega}
\end{align}
with the present-day entropy density $s_0$ and critical density $\rho_{c,0}$, is approximately independent of $m_\chi$.

\begin{figure*}
\includegraphics[width=0.98\textwidth]{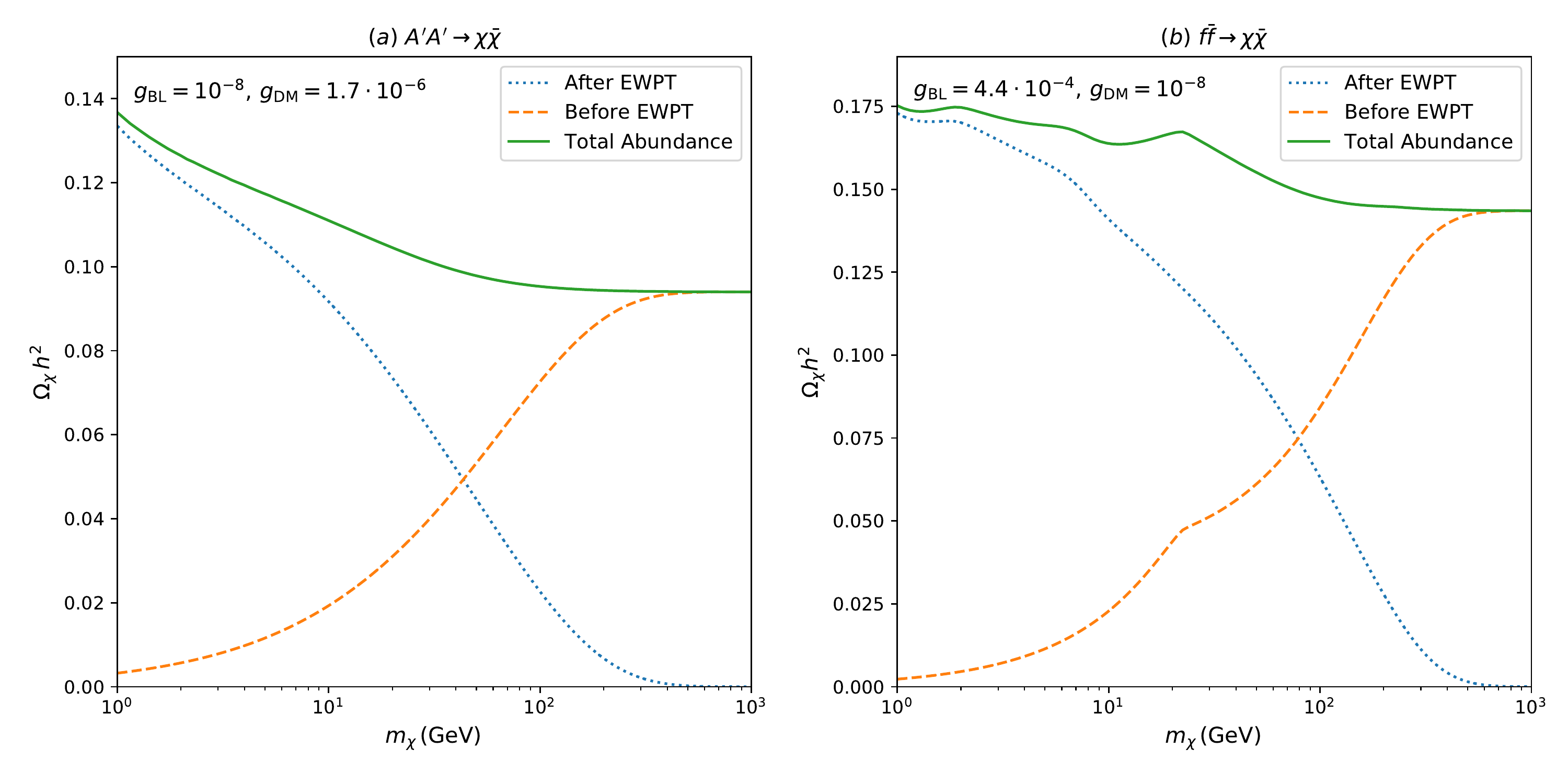}
\caption{\label{fig:Abundance_before_after} The total DM abundance as a function of $m_\chi$ for different limiting cases. Left: $g_\text{DM} \gg g_\mathrm{BL}$ so that production is dominated by $A^\prime$ annihilations. Right: $g_\mathrm{BL} \gg g_\text{DM}$ so that production is dominated by annihilations of SM particles. The orange dashed and blue dotted lines indicate the contribution from temperature above and below the EWPT, respectively.}
\end{figure*}

Refining the rough estimate from above is a difficult task for a number of reasons. First, the particles in the initial state cannot be treated as non-relativistic and therefore need to be described by either Fermi-Dirac or Bose-Einstein statistic. Second, as discussed above, it is important to include thermal corrections to the masses of SM particles as well as to the induced mixing between the $A^\prime$ and the SM photon. Finally, for the case of production from SM particles, we need to make a distinction between times before and after the EWPT~\cite{Heeba:2018wtf}. Most importantly, before the EWPT, the mixing between the dark and the visible photon is replaced by the mixing between the dark photon and the hypercharge gauge boson. At sufficiently high temperatures, DM particles can therefore be produced through \emph{hypercharge plasmon decays}. The difference in the structure of the Lagrangian before and after the EWPT hence makes it necessary to divide the production history into two regimes, $T < T_\mathrm{c} = 164\,\mathrm{GeV}$ and $T> T_\mathrm{c} $.

To take into account all of the effects discussed above, we have implemented the model in a modified version of \texttt{micrOMEGAs\_5}~\cite{Belanger:2018mqt} to solve the Boltzmann equation numerically. The evolution of the DM abundance as a function of temperature is shown in Fig.~\ref{fig:Abundance_temp} for two different values of the DM mass. In both cases we consider $g_\text{DM} = g_\text{BL}$, corresponding to a coupling ratio of $r=1$. Note that for $m_{A^\prime}<1\,\mathrm{GeV}$, the freeze-in production is found to be independent of $m_{A^\prime}$.

We find that the two different production modes discussed above give comparable contributions, although the dark photon channel $A^\prime A^\prime \to \chi\bar{\chi}$ is slightly more efficient. For the case $m_\chi = 30\,\mathrm{GeV}$, most of the DM production happens after the EWPT, whereas for $m_\chi = 300\,\mathrm{GeV}$, the dominant contribution comes from higher temperatures. In the latter case, one can also clearly see a kink in the contribution from SM particles in the initial state at the temperature when hypercharge plasmon decays become kinematically forbidden.

Fig.~\ref{fig:Abundance_before_after} shows the predicted DM abundance as a function of DM mass when considering only dark photons (left) and only SM particles (right) in the initial state. As expected from the rough estimate above, the sum of the contribution from temperatures above the EWPT (orange dashed) and from lower temperatures (blue dotted) gives an abundance that is largely independent of the DM mass. The small wiggles in the right panel result from the fact that for DM masses below 10 GeV the number of relativistic SM fermions that contribute to the freeze-in production is reduced. Neglecting the slight mass dependence, we obtain the approximate expression
\begin{equation}
\Omega_\chi h^2 \approx \Big( 0.16 \, r^{-4} + 0.12 \, r^{-2} \Big) \left(\frac{g_\text{BL}}{2 \cdot 10^{-6}}\right)^4 \; .
\end{equation}

\begin{figure*}
\includegraphics[width=0.98\columnwidth]{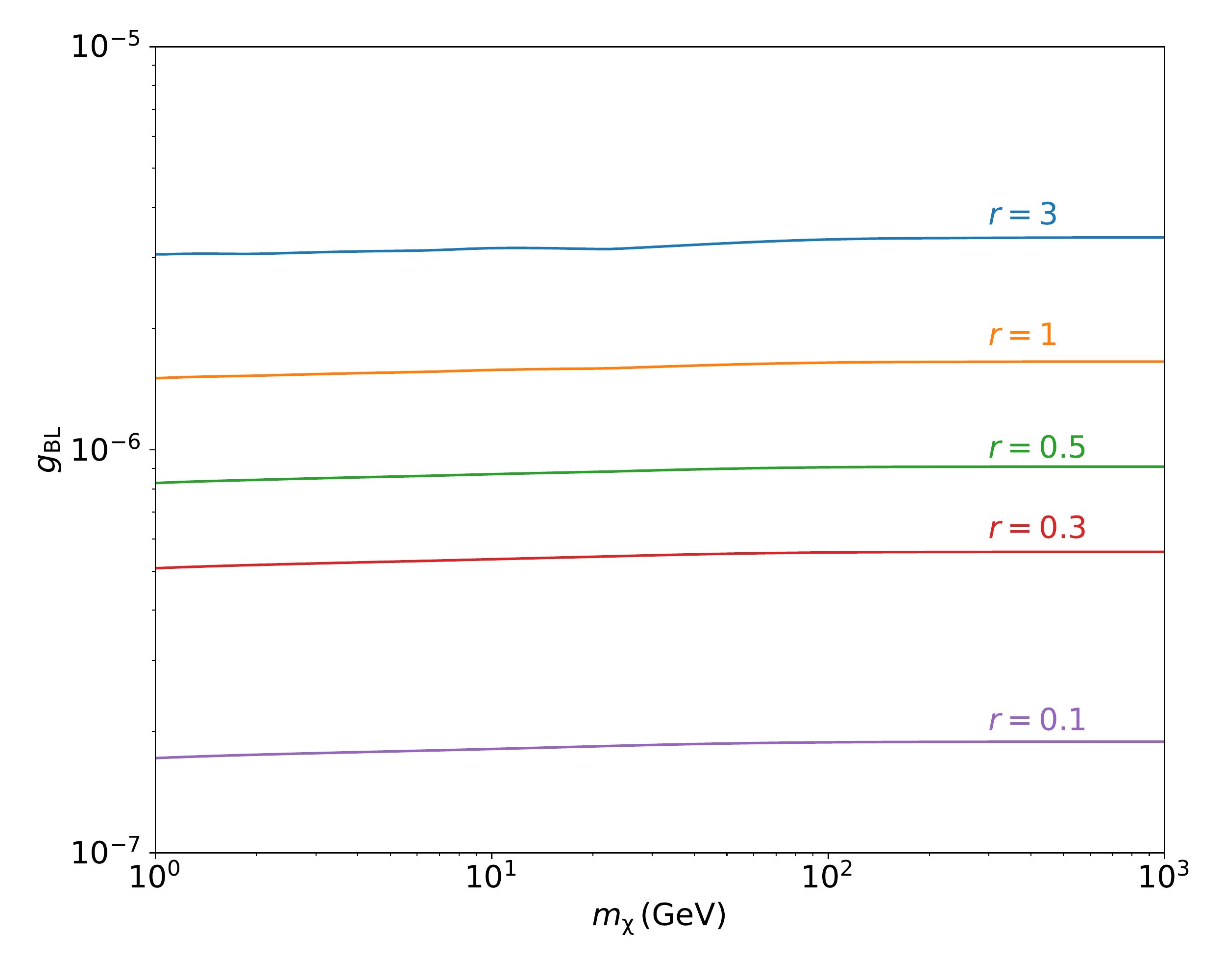}\qquad
\includegraphics[width=0.98\columnwidth]{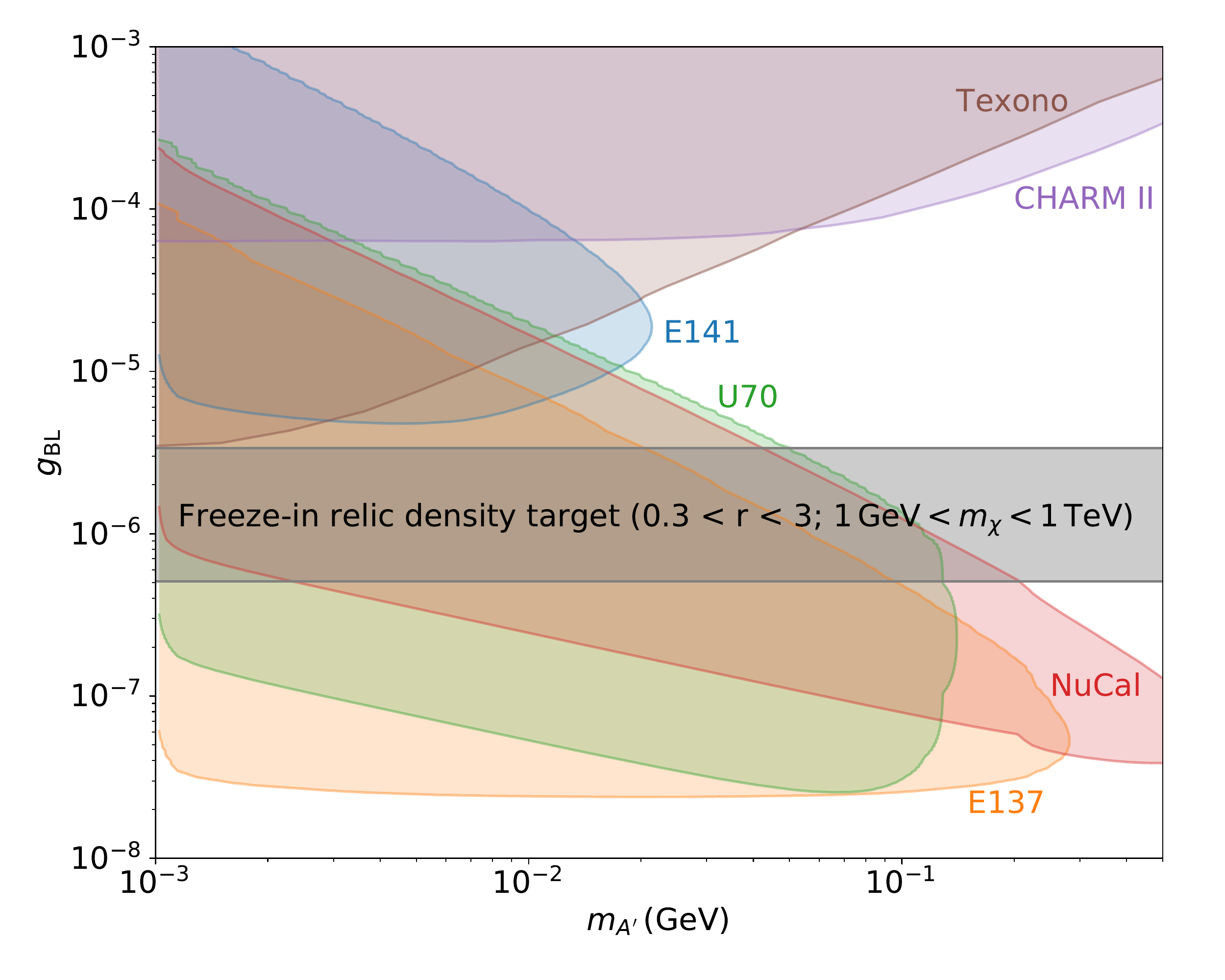}
\caption{\label{fig:coupling_str} Left: Values of $g_\mathrm{BL}$ that give the correct relic abundance for different coupling ratios $r=g_\mathrm{BL}/g_\text{DM}$ as a function of $m_\chi$. Right: the relic density target for $0.3 < r < 3$ and $1 \, \mathrm{GeV} < m_\chi < 1 \, \mathrm{TeV}$ in the $m_{A^\prime}$-$g_\mathrm{BL}$ parameter plane.}
\end{figure*}

Finally, we can use the Planck constraint on DM relic abundance, $\Omega_\chi h^2 = 0.12$, \cite{Aghanim:2018eyx} to determine the coupling strength $g_\mathrm{BL}$ and $g_\text{DM}$ as a function of the DM mass. In the left panel of Fig.~\ref{fig:coupling_str}, we show the required value of $g_\mathrm{BL}$ for different values of the coupling ratio $r$. For the case that production from dark photon dominates ($r \ll 1$), the correct relic abundance is obtained for $g_\text{BL} \sim 10^{-6} r$ (or $g_\text{DM} \sim 10^{-6}$), while for the case $r \gg 1$ (such that SM particles in the initial state give the dominant contribution) we find $g_\mathrm{BL} \sim 10^{-6} \sqrt{r}$ (or $g_\text{BL} \, g_\text{DM} \sim 10^{-12}$).

\subsection*{Relic density targets for accelerator experiments}

As shown above, a new $U(1)_{B-L}$ gauge boson with mass in the MeV range may mediate the interactions responsible for the production of DM particles via the freeze-in mechanism. The required coupling strength is of the order of $g_\mathrm{BL} \sim 10^{-6}$ with some dependence on the assumed $B-L$ charge of the DM particle (parametrised by the coupling ratio $r$) and a weak dependence on $m_\chi$. Intriguingly, dark photons with mass and couplings in this range can be searched for at beam dump, fixed target and collider experiments. Furthermore, the direct coupling to neutrinos in $B-L$ models provides an additional avenue to constrain these models using neutrino experiments such as Texono~\cite{Deniz:2009mu}  (see also Ref.~\cite{Bilmis:2015lja}). For a detailed review of these constraints, we refer the reader to Ref.~\cite{Bauer:2018onh}.

We can translate the results of the relic density calculation from above to the $g_\mathrm{BL}$-$m_{A'}$ parameter plane in order to compare them to the sensitivity of particle physics experiments. For the purpose of this translation it is reasonable to assume that the $B-L$ charge of the DM particle, and hence the coupling ratio $r$ will be of the order of unity. Restricting ourselves to the range $0.3 < r < 3$, we can use the relic density requirement to obtain a narrow range of couplings $g_\mathrm{BL}$ that provide a clear target for accelerator experiments. The resulting band is shown in the right panel of Fig.~\ref{fig:coupling_str} together with existing constraints on dark photons from a $U(1)_{B-L}$ gauge extension (taken from Ref.~\cite{Bauer:2018onh}). In addition to variations in the coupling ratio $r$, the width of the band also reflects the spread of predictions as the DM mass is varied in the range $1\,\mathrm{GeV} < m_\chi < 1\,\mathrm{TeV}$. Since the predicted DM abundance is independent of $m_{A'}$, the band is exactly horizontal.

\section{Probing freeze-in with direct detection}
\label{sec:constraints}

In addition to the general constraints on new gauge bosons shown in Fig.~\ref{fig:coupling_str}, there are important constraints arising from the interactions of the DM particles themselves. Indeed, direct detection experiments are known to provide some of the strongest constraints on models of thermal DM. For non-thermal production via the freeze-in mechanism the required couplings are much smaller, making these models very challenging to probe. However, if the mediator of DM-nucleus scattering is much lighter than the DM particle, direct detection cross-sections can be strongly enhanced, which can potentially compensate for the suppression from small couplings~\cite{Hambye:2018dpi}. In our case the differential DM-nucleus scattering cross section with respect to recoil energy $E_\mathrm{R}$ is given by
\begin{equation}
 \frac{\mathrm{d}\sigma_N}{\mathrm{d}E_\mathrm{R}} = \frac{g_\mathrm{BL}^2 \, g_\text{DM}^2 \, m_N \, A^2 \, F^2(E_\mathrm{R})}{2 \pi \, v^2 (2 \, m_N \, E_\mathrm{R} + m_{A^\prime}^2)^2} \; ,
\end{equation}
where $m_N$ and $A$ denote the mass and mass number of the nucleus, $F(E_\mathrm{R})$ is the nuclear form factor and $v$ is the velocity of the incoming DM particle. 

The smallest recoil energy observable in the XENON1T experiment is $E_\mathrm{th} = 1.1\,\mathrm{keV}$. Hence, for $m_{A'} \ll \sqrt{2 \, m_N \, E_\mathrm{th}} \approx 16 \, \mathrm{MeV}$ the differential event rate will be independent of the mediator mass and benefit from the same enhancement as in the case $m_{A'} \to 0$. For larger mediator masses, on the other hand, direct detection signatures will be suppressed proportional to $m_{A'}^{-4}$.

We furthermore observe that for $r \gg 1$, i.e.\ $g_\mathrm{BL} \gg g_\text{DM}$, the direct detection cross section depends on exactly the same coupling combination as the DM relic abundance. Hence, if we focus on coupling combinations that reproduce the observed DM abundance, the direct detection cross section is a function of only $m_\chi$ and $m_{A^\prime}$. For $r \ll 1$ on the other hand, the DM relic abundance is proportional to $g_\text{DM}^4$, whereas the direct detection cross section is proportional to $g_\text{DM}^4 r^2$. Hence, if $g_\text{DM}$ is determined by the observed DM abundance, event rates in direct detection experiments are suppressed proportional to $r^2$, making this case much harder to probe with direct detection experiments.

\begin{figure}
\centering
\includegraphics[width=0.98\columnwidth]{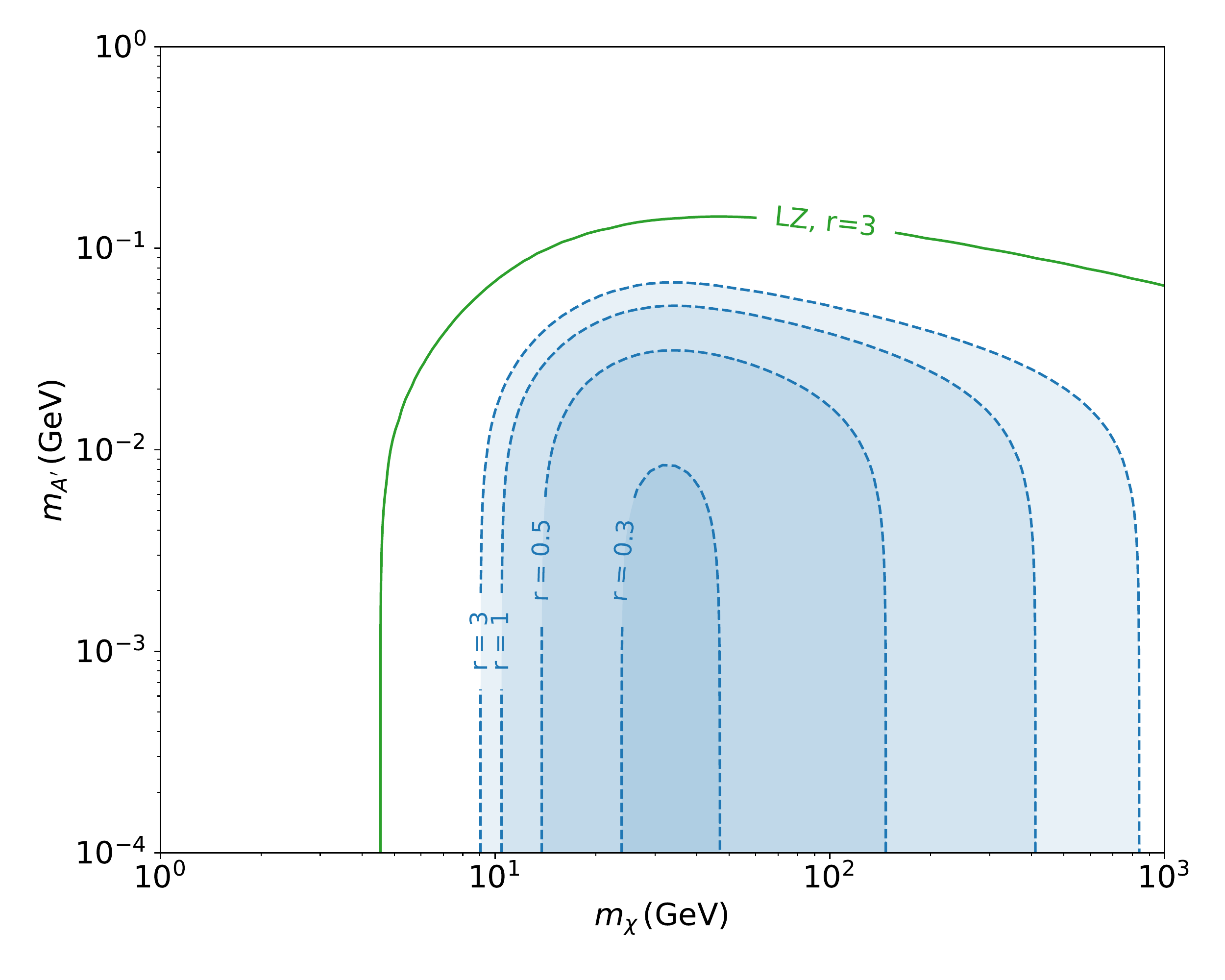}
\caption{\label{fig:DD_lim}Parameter region excluded by the XENON1T experiment for different values of the coupling ratio $r = g_\mathrm{BL}/g_\text{DM}$. For comparison we also show the expected sensitivity of LZ for the case $r = 3$.}
\end{figure}

We use the publicly available code \texttt{DDCalc\_2}~\cite{Athron:2018hpc} and XENON1T data~\cite{Aprile:2018dbl} to place bounds on our model. Our results are shown in Fig.~\ref{fig:DD_lim} for different values of the coupling ratio $r$. At each point in the plot the remaining free parameter (e.g.\ $g_\text{DM}$) has been fixed by the requirement to reproduce the observed DM relic abundance via the freeze-in mechanism. We find that the bounds are strongest for a DM mass of $m_\chi = 30\,\mathrm{GeV}$ and a coupling ratio $r=3$. We do not show any bounds for $r>3$, because they would look very similar to the case with $r = 3$. For decreasing $r$, on the other hand, the bounds become weaker and disappear entirely for $r < 0.3$. We also indicate the expected sensitivity of the LZ experiment~\cite{Akerib:2018lyp} for $r = 3$.

\begin{figure*}
\includegraphics[width=0.98\columnwidth]{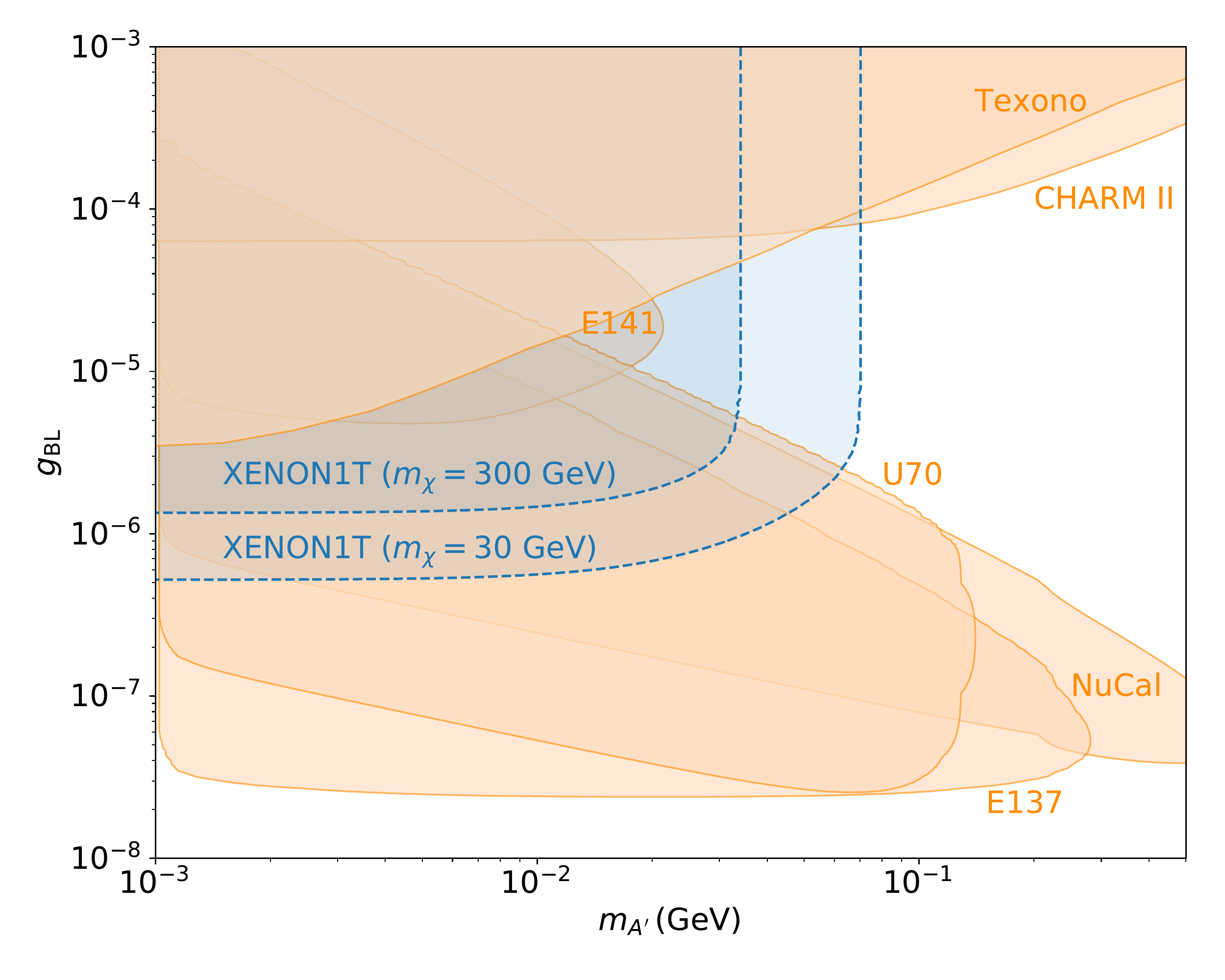}\qquad
\includegraphics[width=0.98\columnwidth]{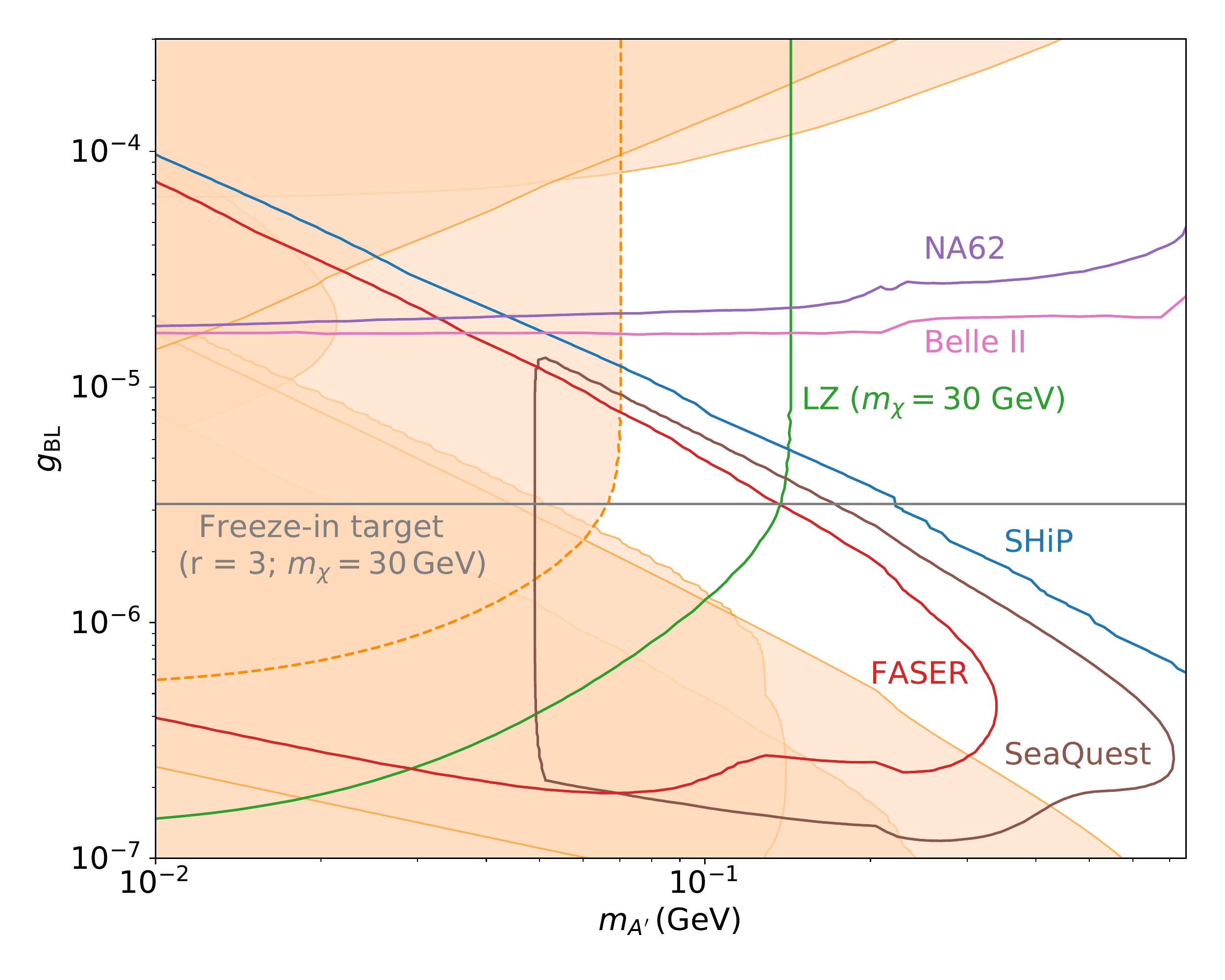}
\caption{\label{fig:master_plot}Comparison of the constraints from direct detection and accelerator experiments for a DM particle produced via the freeze-in mechanism in a $U(1)_{B-L}$ gauge extension of the SM. The direct detection constraints depend on the assumed value of the DM mass and the coupling ratio, $r$, and are shown for $m_\chi = 30\,\mathrm{GeV}$ and $m_\chi = 300\,\mathrm{GeV}$, and $r = 3$. The right panel shows projected sensitivities for various proposed accelerator experiments as well as for the LZ experiment assuming $m_\chi = 30\,\mathrm{GeV}$.}
\end{figure*}

Fig.~\ref{fig:master_plot} presents an alternative way to show the bounds from XENON1T. Instead of fixing the coupling ratio $r$ and varying $m_\chi$ and $m_{A'}$, here we fix the DM mass and vary $m_{A'}$ and $g_\mathrm{BL}$. As before, the coupling $g_\text{DM}$ is determined at each point in the parameter plane by the relic density requirement. As expected, the excluded parameter region becomes independent of the coupling strength for large $g_\mathrm{BL}$ (corresponding to $r \gg 1$) and vanishes for small $g_\mathrm{BL}$ ($r \ll 1$).

The great advantage of this way of presenting direct detection constraints is that they can be directly compared to constraints from accelerator experiments. We find that for the specific model that we consider, direct detection experiments can probe parameter regions that are very difficult to constrain with other kinds of experiments. Of course, these constraints rely on the presence of a DM particle produced via freeze-in and do not apply to general $U(1)_{B-L}$ models. Also, direct detection experiments currently only constrain DM masses between about 10 GeV and 1 TeV. Future experiments like LZ will be able to extend this range of DM masses and also probe larger values of $m_{A'}$. This is illustrated in the right panel of Fig.~\ref{fig:master_plot}, which shows the projected sensitivity of LZ for $m_\chi = 30\,\mathrm{GeV}$ in comparison with future accelerator experiments (taken from Ref.~\cite{Bauer:2018onh}).

\section{Conclusions}
\label{sec:conclusions}

In this work, we have studied the cosmology and phenomenology of a model of DM produced via the freeze-in mechanism with an MeV-scale $U(1)^\prime$ gauge boson. Because of its interactions with the SM plasma and consequent thermalisation, the mediator mixes with the SM hypercharge boson before the EWPT and with the photon after the EWPT. This mixing depends on the effective charge degrees of freedom present in the plasma and therefore differs in different temperature regimes. The DM particle, which we take to be a Dirac fermion with mass in the GeV--TeV range, is then produced by annihilation of the new gauge bosons as well as annihilation of SM particles. Additionally, plasmon and hypercharge plasmon decays provide another non-negligible source of DM production. Which channel dominates depends on the ratio of the mediator-SM coupling $g'$ and the mediator-DM coupling $g_\text{DM}$. We find that the DM relic abundance depends only slightly on the DM mass and is independent of the mediator mass. 

This model can be probed using a number of complementary search strategies. First, an MeV-scale mediator in the coupling range of interest for freeze-in is sufficiently long-lived to be studied with beam dump and fixed target experiments. The resulting limits are independent of the DM mass but constrain the coupling of the mediator to the SM. Second, since the mediator mass can be comparable to the typical momentum transfer in direct detection experiments, direct detection event rates are enhanced and may be observable despite the small freeze-in couplings.

The constraints on the effective DM-SM coupling $g' \times g_\mathrm{DM}$ obtained in this way can be compared to the calculation of the DM relic abundance, which probes a similar coupling combination. One then obtains constraints on the DM and mediator masses, which are particularly strong when $g' > g_\mathrm{DM}$ and weaken in the opposite regime. For a given DM mass, it is possible to represent these constraints in the $g'$-$m_{A^\prime}$ parameter plane in order to directly compare with bounds from accelerator experiments. Our findings are summarised in Fig. \ref{fig:master_plot}, which demonstrates the promising potential of upcoming direct detection and accelerator experiments to probe large parts of the parameter space of this model compatible with the bounds from current searches.

The coupling structure of the new $U(1)^\prime$ provides our model with additional  freedom. Here, we have focused on the case of gauged baryon minus lepton number, which has a number of attractive features such as being anomaly-free and providing a potential explanation of the small masses of left-handed neutrinos via the see-saw mechanism. Our calculations for the mass mixing between the different gauge bosons can however be directly applied to an arbitrary $U(1)^\prime$ extension provided the associated gauge boson thermalises with the SM plasma. For coupling ratios $g' / g_\text{DM}$ of order unity the freeze-in calculation of the relic abundance will only change slightly for different $U(1)^\prime$ extensions. One can then map $g_\mathrm{BL}$ and $g_\mathrm{DM}$ to equivalent portal couplings in other $U(1)^\prime$ models. Of course, the accelerator and direct detection constraints for these models may change substantially and need to be worked out individually (see e.g.\ Refs.~\cite{Bauer:2018onh, Foldenauer:2018zrz}). This remains an interesting question to be examined and is left for future work. 

\vfill

\noindent 
{\bf Acknowledgements.} We thank Julian Heeck for helpful comments on the manuscript, Fatih Ertas, Annika Reinert, Kai Schmidt-Hoberg, Thomas Schwetz and Bryan Zaldivar for useful discussions, Patrick Foldenauer for providing the data points from Ref.~\cite{Bauer:2018onh} and Alexander Pukhov for help with~\texttt{micrOMEGAs\_5}. We also thank the Erwin Schroedinger International Institute for hospitality while this work was completed. This  work  is funded  by  the  Deutsche Forschungsgemeinschaft (DFG) through the  Emmy Noether Grant No.\ KA 4662/1-1 and the Collaborative Research Center TRR 257 ``Particle Physics Phenomenology after the Higgs Discovery''.

\end{document}